\documentclass[12pt]{article}
\usepackage{epsf}
\usepackage{cite}

\usepackage{amsmath}
\usepackage{amsfonts}
\usepackage{amssymb}
\def\be{\begin{equation}}
\def\ee{\end{equation}}
\def\beq{\begin{eqnarray}}
\def\eeq{\end{eqnarray}}

\def\cm1{C^{-1}_{ij}}
\def\fnl{{f_{NL}}}

\def\s2{\sigma^2}

\def\fnlloc{f_{\rm NL}}
\def\tildefnlloc{\tilde{f}_{\rm NL}}

\def\stheta{\sum_{\vec{\theta}}}
\def\sl1l2{\sum_{l_1l_2}}
\def\Npix{{\rm N}_{\rm pix}}
\def\fnl{f_{\rm NL}}
\def\cale{{\cal E}}
\def\lmin{l_{\rm min}}

\def\calk{{\cal K}}

\begin{document}

\begin{flushright} {\footnotesize HUTP-06/A0016}\\ {\footnotesize MIT-CTP 3737}\\  {\footnotesize IC/2006/028} \end{flushright}

\begin{center}
{\Large \bf Estimators for local non-Gaussianities} \\[.5cm]

{\large Paolo Creminelli$^{\rm a}$, Leonardo Senatore$^{\rm b}$ \\
[.15cm] and   Matias Zaldarriaga$^{\rm c,d}$}
\\[0.5cm]

{\small \textit{$^{\rm a}$ Abdus Salam International Center for Theoretical Physics, \\
Strada Costiera 11, 34014 Trieste, Italy}}

\vspace{.2cm}

{\small
\textit{$^{\rm b}$ Center for Theoretical Physics, \\
Massachusetts Institute of Technology, Cambridge, MA 02139, USA }}

\vspace{.2cm}
{\small
\textit{$^{\rm c}$ Jefferson Physical Laboratory, \\
Harvard University, Cambridge, MA 02138, USA}}

\vspace{.2cm}
{\small
\textit{$^{\rm d}$ Center for Astrophysics, \\
Harvard University, Cambridge, MA 02138, USA }}
\end{center}

\hrule \vspace{0.3cm}
{\small  \noindent \textbf{Abstract} \\[0.3cm]
\noindent
We study the Likelihood function of data given $\fnl$ for the
so-called local type of non-Gaussianity. In this case the curvature
perturbation is a non-linear function, local in real space, of a
Gaussian random field. We compute the Cramer-Rao bound for $\fnl$ and
show that for small values of $\fnl$ the 3-point function estimator
saturates the bound and is equivalent to calculating the full
Likelihood of the data. However, for sufficiently large $\fnl$, the
naive 3-point function estimator has a much larger variance than
previously thought. In the limit in which the departure from
Gaussianity is detected with high confidence, error bars on $\fnl$
only decrease as $1/\ln \Npix$ rather than $\Npix^{-1/2}$ as the size
of the data set increases. We identify the physical origin of this
behavior  and explain why it only affects the local type of
non-Gaussianity, where the contribution of the first multipoles is
always relevant. We find a simple improvement to the 3-point function
estimator that makes the square root of its  variance decrease as
$\Npix^{-1/2}$ even for large $\fnl$, asymptotically approaching the
Cramer-Rao bound. We show that using the modified estimator is
practically equivalent to computing the full Likelihood of $\fnl$
given the data.  Thus other statistics of the data, such as the
4-point function and Minkowski functionals, contain no additional
information on $\fnl$. In particular, we  explicitly show that the
recent claims about the relevance of the 4-point function are not
correct. By direct inspection of the Likelihood, we show that the data
do not contain enough information for any statistic to be able to
constrain higher order terms in the relation between the Gaussian
field and the curvature perturbation, unless these are orders of
magnitude larger than the size suggested by the current limits on
$\fnl$. As our main focus is the scaling with $\Npix$ of
the various quantities, calculations are done in flat sky approximation
and without the radiation transfer function.

\vspace{0.5cm}  \hrule

\section{Introduction}
In single field slow-roll inflation the level of non-Gaussianity is sharply predicted and very small, less than $10^{-6}$ \cite{Maldacena:2002vr,Acquaviva:2002ud}. This is quite far from the present experimental sensitivity and probably not attainable with either CMB observations or galaxy redshift surveys. As a result, deviations from a purely Gaussian statistics of density perturbations, if observed, could provide important constraints on models of early cosmology, forcing us to abandon the single-field slow-roll paradigm. 

Of course there are many ways in which a signal could be ``non-Gaussian". Given a data set, such as the WMAP maps, there are two possible ways to proceed. One could calculate all kinds of statistics of the data and compare the results with the expectation for a Gaussian field searching for anomalies. This is a fine strategy as long as one adjusts the significance of the result to account for the number of possible deviations that have been explored. There are several anomalies in the WMAP data reported in the literature  that have been found in this way (see for example \cite{Land:2005ad,Cruz:2006fy,McEwen:2006yc}). Unfortunately their significance is hard to assess and as a result one is not sure how seriously to take them. 

The second approach is to think about the possible physical mechanisms that can lead to non-Gaussianities and search for their particular signatures. In the context of primordial effects one should investigate what types of non-Gaussianity can plausibly  be produced in various inflationary models. This approach is further bolstered by the fact that at least at the level of the 3-point function, primordial signals seem to fall into two definite classes. Thus there are only two different signatures one has to look for. 

The analysis of inflationary models that go beyond the single field slow-roll class has identified several examples with a relatively high level of non-Gaussianity, within reach of present or forthcoming experiments. 
For nearly Gaussian fluctuations, the quantity most 
sensitive to departures from perfect Gaussianity is the 3-point correlation function. 
In general, each model will give a different correlation 
between the Newtonian potential modes\footnote{Even with perfectly Gaussian primordial fluctuations, the observables, {\em e.g.} the temperature anisotropy, will not be perfectly Gaussian as a consequence of the non-linear relation between primordial perturbations and what we will eventually observe. These effects are usually of order $10^{-5}$ (see for example \cite{Creminelli:2004pv,Bartolo:2006cu}) and thus beyond (but not much) present sensitivity. In the following we will disregard these contributions.}:
\be
\langle \Phi({\bf k}_1) \Phi({\bf k}_2) \Phi({\bf k}_3) \rangle = (2 \pi)^3 \delta^3
\big({\bf k}_1 + {\bf k}_2 + {\bf k}_3 \big)
F( k_1,  k_2 ,  k_3) \;.
\ee
The function $F$ describes the correlation as a function of the triangle shape in momentum space. 

The predictions for the function $F$ in different models divide quite sharply into two qualitatively different classes as a consequence of qualitatively different ways of producing correlations among modes \cite{Babich:2004gb}. The first possibility is that the source of density perturbations is not the inflaton but a second light scalar field
$\sigma$. In this case non-Gaussianities are generated by the non-linear relation between the fluctuation $\delta\sigma$ of this field 
and the final perturbation $\Phi$ we observe. This non-linearity is {\em local} as it acts
when the modes are much outside the horizon; schematically we have  
$\Phi ({\bf x}) = g({\bf x}) + \fnl (g^2({\bf x}) - \langle g^2\rangle) + \ldots$, where $g$ is a Gaussian random field.  
The quadratic piece introduces a 3-point function 
for $\Phi$ of the form
\be
\label{eq:local}
F( k_1,  k_2 ,  k_3) = \fnl \cdot 2 \Delta_\Phi^2 \cdot \left(\frac1{k_1^3 k_2^3} + \frac1{k_1^3 k_3^3} + 
\frac1{k_2^3 k_3^3}\right) \;,
\ee
where $\Delta_\Phi$ is the power spectrum normalization,
$\langle \Phi({\bf k}_1) \Phi({\bf k}_2) \rangle = (2\pi)^3 \delta^3\big({\bf k}_1 + {\bf k}_2\big)
\Delta_\Phi \cdot k_1^{-3}$, which has been taken as exactly scale invariant.
Examples of this mechanism are the curvaton scenario \cite{Lyth:2002my} and the variable decay width model \cite{Zaldarriaga:2003my}, which naturally give rise to $\fnl$ greater than 10 and 5, respectively. Various subtleties in estimating the size of this type of non-Gaussianity will be the focus of this paper.

The second class of models are single field models with a non-minimal Lagrangian, where the correlation among modes is created by higher derivative operators \cite{Creminelli:2003iq,Arkani-Hamed:2003uz,Alishahiha:2004eh,Senatore:2004rj,Chen:2006nt}. In this case, the correlation 
is strong among modes with comparable wavelength and it decays when we take one of $k$'s to zero
keeping the other two fixed. Although different models of this class give a different function $F$, all these functions are qualitatively very similar. We will call this kind of functions {\em equilateral} because the signal is maximal for equilateral configurations in Fourier space, whereas for the local form (\ref{eq:local}) the most relevant configurations are the {\em squeezed} triangles with one side much smaller than the others.  We will not discuss the {\em equilateral} type of non-Gaussianity in this paper too much. We will just point out that the effects studied in this paper do not apply in that case so that the situation is much simpler.


The strongest constraint on $\fnl$ comes from analyzing  the 3-point function of the WMAP  data set. WMAP is the best available data set because  it has the largest number of pixels measured with good signal-to-noise. From the first year data the constraint is \cite{Creminelli:2005hu}:
\be
-27 < \fnl < 121  \quad {\rm at} \;95\% \;{\rm C.L.}
\ee
This constraint is better than that obtained by the WMAP collaboration both using the one year WMAP data \cite{Komatsu:2003fd} and the three year ones \cite{Spergel:2006hy}. This is so because the WMAP team used a non-optimal estimator which did not adequately treat the effect of anisotropic noise, as already noted in \cite{Komatsu:2003fd,Komatsu:2003iq}. In  \cite{Creminelli:2005hu}, we showed that the effect of the anisotropic noise can be substantially reduced with the addition of a linear piece to the estimator. Always in \cite{Creminelli:2005hu} we also constrained the level of the {\it equilateral} 3-point function. For both types, the departures from Gaussianity still allowed by the data are at the same level. 


Given the interest in constraining the level of non-Gaussianity, one may wonder if a statistic other than the 3-point function might extract more information about $\fnl$. There are various contradictory, or at least apparently contradictory,  answers to this question in the literature. On the one hand in \cite{Babich:2005en} and \cite{Creminelli:2005hu} it is argued that the 3-point function saturates the Cramer-Rao bound up to terms of order  $\fnl A^{1/2}$, where $A$ is the square of the amplitude of curvature perturbations: $A^{1/2} \sim 10^{-5}$. On the other hand calculations of the signal to noise in the 4-point function by \cite{Okamoto:2002ik} and \cite{Kogo:2006kh} point to a different conclusion. These papers claim that, even though in the limit of $\fnl \rightarrow 0$ the signal to noise ratio of the 4-point function is negligible, it grows more rapidly with the number of pixels in the data set than for the 3-point function. As a result for values of $\fnl$ rather small, say around $\sim 50$ for an experiment like Planck, the signal to noise in the 4-point function is larger than for the 3-point function and stronger constraints on $\fnl$ could be placed by studying the 4-point function. Of course this result is puzzling. One is immediately drawn to the question, what about the 5-point function? And why not the 11-point function? Applying the same arguments as in  \cite{Okamoto:2002ik} and \cite{Kogo:2006kh} would show that the signal to noise ratio becomes larger the higher the n-point function considered. Clearly there is a contradiction. 


It is the aim of this paper to clarify this contradiction. We will show that both calculations have missed an interesting subtlety of the local type of non-Gaussianity in the case of scale invariant, or nearly scale invariant, spectrum of primordial perturbations. As a result, the calculation of the noise of various estimators (including the 3-point function) for finite $\fnl$ is missing some relevant term. Some of the terms that are naively down by powers of $\fnl A^{1/2}$ are actually much larger, being enhanced by $\Npix$. The growth in the signal to noise for high  $\fnl$ seen in the above papers is fictitious. We will show that the same subtlety creeps into the calculation of \cite{Babich:2005en} and thus the 3-point function estimator considered there also does not saturate the Cramer-Rao bound for large $\fnl$. We want to stress that even though what was missed was a rather subtle point, it has potentially large consequences on the signal to noise of the estimators previously considered. For example when one is in the regime of large signal to noise, the error bars on $\fnl$ from the 3-point function decrease as $1/\ln \Npix$ rather than $\Npix^{-1/2}$ . The reader at this point should not panic, we will show that the Cramer-Rao bound in this regime  still scales as $\Npix^{-1/2}$ and that it is rather straightforward to extract all of this information from the data either by calculating the full Likelihood or slightly tweaking the 3-point function estimator. 


What is the missing subtlety? To understand it, it is best to recall what is the main effect of the local non-Gaussianity: it correlates large and small scales.  In the 3-point function, a long wavelength mode modulates the amplitude of all the short wavelengths by the same amount, regardless of the wavelength of the short mode.  Furthermore, in \cite{Babich:2004gb} we showed that most of the signal in the 3-point function is coming from squeezed triangular configuration in Fourier space.  More importantly for this discussion, if one considers the signal to noise as a function of the wavenumber of the long wavelength mode $k_L$,  one gets an equal amount of  information from every logarithmic interval in $k_L$. This in fact is the source of the problem. The smallest $k_L$ in the survey are by definition the ones with the largest cosmic variance as there are the least of them in the survey. As one increases the resolution of the survey the contribution to the signal to noise from the long wavelengths only decreases logarithmically and thus the large cosmic variance of  the long modes translates into large variances in the estimators of $\fnl$.


In fact in \cite{Creminelli:2005hu} indications of the importance of the long wavelength modes were emphasizes in the context of the effect of anisotropic noise on the estimator of the 3-point function. The noise in the map is anisotropic because WMAP spent different amounts of time observing each pixel on the sky.  As a result the level of small scale power, for large multipoles where the noise becomes important,  varies across the sky. This map of small scale power can randomly align with the particular large scale mode giving a spurious $\fnl$ signal. Of course on average this effect is zero as there is no intrinsic correlation between the map of observing time per pixel and the large scale temperature.  However for a particular realization, some modes  will be correlated (spurious positive $\fnl$)  and  others anti-correlated (spurious negative $\fnl$).   The contribution to the signal from the long wavelength modes will not add exactly to zero, as we have few of them in the survey.  The random left over spurious signal effectively increases the variance of the estimator. This effect was noted in the WMAP team analysis \cite{Komatsu:2003fd,Komatsu:2003iq}, where the constraint on $\fnl$  got worse as they increased the size of the data set by including more of the small scales. In \cite{Creminelli:2005hu}  the estimator was improved by including a linear piece which substantially reduces the effect  allowing us to get better constraints. This effect is the reason why the 3-year data analysis by the WMAP  team \cite{Spergel:2006hy} did not appreciably improve the limits.


To clarify the situation we will study the full Likelihood of the data given $\fnl$. We will keep careful track of enhanced terms and thus do a consistent expansion in $\fnl$. We will then calculate the Cramer-Rao bound, extending the results of \cite{Babich:2005en} to non-zero $\fnl$. We will show how the additional terms in the variance of the 3-point function estimator make it become sub-optimal. This can be easily fixed using an improved estimator which asymptotically saturates the Cramer-Rao bound. The use of this estimator is equivalent to the full Likelihood of the data.

The fact that the improved 3-point function estimator is equivalent to the full Likelihood of the data, implies that there is no additional information in the 4-point function. We will also show this explicitly,
illustrating how at best the 4-point function is equivalent to the 3-point function.
No other statistic such as Minkowski functionals, various wavelet based statistics and other esoteric constructions are worth trying to constrain $\fnl$. {\em None can be better than the 3-point function.} This is true up to corrections of order $\fnl A^{1/2} \lesssim 10^{-3}$.


 The apparent large signal to noise in the 4-point function led to the suggestion \cite{Okamoto:2002ik,Kogo:2006kh} that the 4-point function could even be sensitive to higher order terms in the relation between $\Phi$ and $\delta\sigma$: 
$\Phi ({\bf x}) =  g({\bf x}) + \fnl (g^2({\bf x}) - \langle g^2\rangle) +  \fnl^2 \alpha  g^3({\bf x})  \ldots$ The claim was that the 4-point function could constrain the real parameter $\alpha$.  Of course the third term is a minuscule correction to the first two, even for the largest allowed values of  $\fnl$. Thus it is  difficult to understand how one could be sensitive to it. Again the missing terms in the variance of the estimator were responsible for the apparent sensitivity to $\alpha$. Using the full Likelihood we will show that for any realistic experiment there is in fact not enough information about $\alpha$ in the data to constrain it, unless $\alpha > 1/  (\fnl A^{1/2}) \gtrsim 10^3$. 

Finally we will also show that for a realistic experiment where  $\ln
\Npix$ is large, the value of $\fnl$ for which improving  the naive
3-point function estimator is important is rather large. One should
start worrying about it once there is a many $\sigma$ detection of
$\fnl$.  As a result our improved estimator will probably be only of
academic interest. Our paper mainly provides clarification of various
misconceptions in the literature. Given this and to reduce the length
of our equations, we will work in the flat sky approximation and
neglect the CMB transfer functions, directly working with a
2-dimensional random field with a local non-Gaussianity. Expressions
for the 3 and 4-point functions including the radiation transfer
function and with spherical geometry can be found for instance in \cite{Komatsu:2002db, Kogo:2006kh}.

We want to stress\footnote{We thank the unknown referees for
  correspondence about this point.} that our approximation captures the
qualitative features of the real problem, like the dependence of the
various expressions on the number
of data $N_{\rm pix}$. On the other hand one should not trust the
numerical factors that we will find, because they would be changed in
a complete treatment, where projection effects from 3 to 2 dimensions
are taken into account together with the full radiation transfer function.
The reason why the qualitative features are captured by our
approximation is that, although a given mode on the sky
receives contribution from a range of different 3D wavelengths, this
effect is limited to an interval $\Delta k \sim k$. This implies that
a squeezed configuration of the 3-point function of primordial
perturbations $k_1 \ll k_2, k_3$ (which, as we will see, are the
only configurations material to our conclusions) maps to a
correlation among different multipoles with $l_1 \ll l_2, l_3$. In
other words although the 3-point function of the 2-dimensional
temperature map is not exactly of the local form, it still roughly
behaves as $\langle a_{\vec l_1} a_{\vec l_2}  a_{\vec l_3}\rangle \propto l_1^{-2} l_2^{-2}
+ {\rm perm.}$, which is the 2-d analogue of eq.~(\ref{eq:local}),
with additional modulations induced by the transfer
function. As we will discuss, our results just depend on the $l_1^{-2}$ dependence for
$l_1 \to 0$ (compared for instance with $l_1^{\, 0}$ which is typical of
other shapes of non-Gaussianity), so that it is enough to stick to an exactly
local non-Gaussianity in 2-dimensions. Also the flat sky approximation
will contribute to change the numerical factors, but leave unaltered
the qualitative features. 

We also point out that even though we  do all our calculations in 2 dimensions, as relevant to the CMB, our conclusions are equally valid for 3 dimensional surveys, such as galaxy surveys or future 21 cm observations.

\section{Subtleties of the $\fnl$ expansion \label{subtleties}}

The aim of this section is to explicitly show the presence of terms in the variance of the 3-point function  estimator of  $\fnl$ that are enhanced by factors of $\Npix$ and thus contribute significantly even though they are naively suppressed by  $\fnl A^{1/2}$. We start by introducing our notation and reproducing previous calculations of the variance of the estimator. We then identify the terms that had been previously missed and give a rule of thumb to easily determine when they are important.  We will show that in practice the enhanced terms do not correct current upper limits and that they will only become important after a very high signal to noise detection of $\fnl$.

\paragraph{Notation.} We work in the flat sky approximation, neglect the transfer
function, and assume that the error is dominated by cosmic
variance. In this paper we are mostly interested in the scaling
properties of the estimators for the non-Gaussianities used for
example in
\cite{Komatsu:2003fd,Babich:2005en,Creminelli:2005hu,Okamoto:2002ik,Kogo:2006kh,Komatsu:2001rj,Babich:2004yc}.
These properties are not modified by these
approximations\footnote{This is confirmed by the
fact we are able to recover the same scaling properties found in
\cite{Babich:2005en,Okamoto:2002ik,Kogo:2006kh,Babich:2004yc,Komatsu:2001rj} where these
approximations were not used.}, while on the other hand their use
makes the presence of some physical effects much clearer, as we will
later see. Let us briefly set up our conventions. For the Fourier
transform we have:
\begin{equation}
\Phi_{\vec{l}}=\frac{\Omega}{\Npix} \stheta e^{-i
\vec{\theta}\cdot\vec{l}} \Phi_{\vec\theta} \ ,
\end{equation}
where $\Omega$ and $\Npix$ are respectively the angular size
and the number of pixels of the sky survey. It is immediate to
obtain from this the continuum limit
$\Phi_{\vec{l}}=\frac{\Omega}{\Npix} \stheta e^{-i
\vec{\theta}\cdot\vec{l}} \Phi_{\vec\theta}\simeq \int
d^2\theta\, e^{-i \vec{\theta}\cdot\vec{l}} \Phi_{\vec\theta}$. We
also have:
\begin{equation}
\Phi_{\vec{\theta}}= \frac{1}{\Omega}\sum_{\vec{l}} e^{i
\vec{\theta}\cdot\vec{l}} \Phi_{\vec{l}}\simeq \int
\frac{d^2l}{(2\pi)^2}\, e^{i \vec{\theta}\cdot\vec{l}}
\Phi_{\vec{l}} \ ,
\end{equation}
and the useful relations:
\begin{equation}
\sum_{\vec{l}}
e^{i\vec{l}\cdot(\vec{\theta}_1-\vec{\theta}_2)}=\Npix \delta_{\vec{\theta}_1,\vec{\theta}_2} \ ,\ \ \ \ \ \ \ \ \
\stheta e^{-i\vec{\theta}\cdot(\vec{l}_1-\vec{l}_2)}=\Npix\delta_{\vec{l}_1,\vec{l}_2} .
\end{equation}

We are only interested in local non-Gaussianities as the effect we will discuss does not apply to other types. In that case, the observed field  $\Phi_{\vec{\theta}}$ is given by a
non linear function of a Gaussian field $g_{\vec{\theta}}$ which
is local in real space:
\begin{equation}
\Phi_{\vec{\theta}}=f(g_{\vec{\theta}})=g_{\vec{\theta}}+\fnlloc\left(g_{\vec{\theta}}^2-\sigma^2\right).
 \label{localnongaussdef}
\end{equation}
We will call this field temperature although our results apply to other measures, not just the CMB temperature. In Fourier space the local relation reads:
\begin{equation}
\Phi_{\vec{l}}=\left(f(g)\right)_{\vec{l}}=g_{\vec{l}}+\fnlloc\left(\left(g\circ
g \right)_{\vec l}-\sigma^2 \Omega \delta_{\vec{l},\,0}\right),
\end{equation}
where we have defined $\left(g \circ g
\right)_{\vec{l}}=\frac{1}{\Omega}\sum_{\vec{k}} g_{\vec{l}-\vec{k}} g_{\vec{k}}$.
We will explicitly address later the case of possible higher order
corrections in $\fnlloc$ to these definitions. The covariance
matrix is defined as
\begin{equation}
\langle
g_{\vec{l}_1}g_{\vec{l}_2}\rangle=C_{\vec{l}_1\vec{l}_2}=
C_{l_1}  \Omega\;\delta_{\vec{l}_1,-\vec{l}_2} ,
\end{equation}
 where $l^2=\vec{l}\cdot \vec{l}$, and
$C_l=2\pi A/l^2$, which then implies that
\begin{equation}
\sigma^2=\langle
g_{\vec{\theta}_i}g_{\vec{\theta}_i}\rangle=\frac{2\pi
A}{\Omega}\sum_{\vec{l}}\frac{1}{l^2}\simeq
\frac{A}{2}\ln\Npix ,
\end{equation}
where in the last passage we have used the continuum limit, and
the fact that $\Npix\simeq \Omega\, l^2_{\rm max}/(4\pi)$, with
$l_{\rm max}$ the maximum of the observed $l$s. From here on, in
order to simplify the notation, we will remove the vector symbol
from $\vec{l}$ and $\vec{\theta}$ in all the mathematical
expressions when the meaning and the distinction from the modulus
$l=|\vec{l}|$ and $\theta=|\vec{\theta}|$ is clear from the
context.


\paragraph{ Previous results: the missing enhanced terms. }
In \cite{Creminelli:2005hu,Komatsu:2003fd} the analysis for the
non-Gaussianities of the local kind was performed using a
trilinear estimator with signal-to-noise weighting. In the limit of flat sky, unit transfer function,
and isotropic noise it reduces to:
\begin{equation}
\cale=\frac{1}{N }\sum_{l}\frac{1}{\Omega\, C_{l}}
\Phi_{l}\chi_{-l} \,, \label{estimator}
\end{equation}
where we have defined the field
$\chi_{l}=(\Phi\circ\Phi)_{l}-\Omega\, \sigma^2 \delta_{l,0}$ , and where 
we consider only non degenerate configurations with all the $\Phi$s taken with $l\neq0$ . The normalization
\begin{equation}
\label{eq:N}
N=\sum_{l}\frac{1}{\Omega\, C_{l}}
\langle\Phi_{l}\chi_{-l}\rangle_{1} \simeq8\Npix\sigma^2 \;,
\end{equation}
with the subscript $_1$ meaning that the expectation value is
taken with $\fnlloc=1$, has been chosen so that the estimator is unbiased, 
$\langle\cale\rangle=\fnlloc$ (\footnote{Some of the expressions, like eq.~(\ref{eq:N}) above, when expressed in terms of $\Npix$ and $\sigma$, will slightly depend on the geometry of the survey which changes the boundary of the domain of integration in Fourier space. Also we will have similar corrections going from flat to full sky. However these effects do not change significantly our results.}). The definition in
eq.~(\ref{localnongaussdef}) tells us that the temperature field in the
sky $\Phi_{\theta}$ is to a good approximation a Gaussian field,
with a small non-Gaussian correction of order $\fnlloc \Phi\sim
\fnlloc A^{1/2}\lesssim 10^{-3}$. So one is tempted to expect that
higher order corrections in $\fnlloc$ in the various expressions
are suppressed
with respect to the leading terms by powers of $\fnlloc A^{1/2}$
and thus irrelevant. For example, the variance of the
estimator in eq.~(\ref{estimator}) starts with a piece which is of
zeroth order in $\fnlloc$, and which gives:
\begin{equation}\label{var0}
\langle\Delta \cale^2\rangle_{\fnlloc=0}=\frac{1}{\Omega^2N^2}\sum_{l
l'}\frac1{C_l C_{l'}}\langle\Phi_{l} \chi_{-l}\Phi_{l'}
\chi_{-l'}\rangle_{\fnlloc=0}=\frac{1}{N} = \frac{1}{8\Npix\sigma^2} \;,
\end{equation}
where $\Delta\cale = \cale-\langle\cale\rangle$ \footnote{Notice that the variance of the
estimator scales faster than the naive $1/\Npix$ by a factor of $\ln\Npix$. This behaviour is typical of 
non-Gaussianities of the local kind, where the signal comes from the correlation of the modes of all
different scales.}.  One would naively assumes that this is the dominant term in the
variance, with small corrections of order $\fnlloc A^{1/2}$, which
should contribute at most at order $10^{-3}$ given the current
bound on $\fnl$. This is what was assumed for example in
\cite{Komatsu:2003fd,Creminelli:2005hu,Babich:2005en}, where the
estimator (\ref{estimator}) was in fact found minimizing the
variance at zeroth order in $\fnlloc$ among all trilinear
estimators. However, as we will soon see, for the case of local
non-Gaussianities, and only for them, there is another parameter
which enters into the expansion: $\Npix$. We will see in fact that
in certain expressions such as the variance of the estimator above for
example, there are terms that although suppressed by powers
of $\fnlloc A^{1/2}$, are enhanced by powers of $\Npix$, and so,
depending on the real value of $\fnlloc$, they might need to be
taken into account.

In  order to verify that this is actually the case, and 
to understand the implications of this fact, let us sketch the computation of the variance of the estimator $\cale$ keeping higher order terms in $\fnlloc$. The variance of $\cale$ will
involve the computation of a 6-point function, which will split
in the sum of the product of several different combinations of
connected $n$-point functions, i.e.~the product of three 2-point
functions, of two 3-point functions, and of a 4-point function and
a 2-point function. Concentrating on the last kind of
contribution, we will have terms like:
\begin{eqnarray}
&&\langle\Delta\cale^2\rangle
\supset\frac{1}{N^2\,\Omega^4}\sum_{l_1l_2\tilde{l}_1\tilde{l}_2}\frac{1}{C_{l_2}C_{\tilde
l_2}}\langle\Phi_{l_1}\Phi_{\tilde{l}_1}\rangle_c\langle\Phi_{l_2}\Phi_{-l_1-l_2}
\Phi_{\tilde{l}_2} \Phi_{-\tilde{l}_1-\tilde{l}_2}\rangle_c \;,
\label{exvar}
\end{eqnarray}
where $\langle\Phi_{l_1}\Phi_{l_2}\cdots\Phi_{l_n}\rangle_c$ stays
for the connected $n$-point function. Now, apart from numerical factors, one of the terms in
the expansion of the connected 4-point function reads:
\begin{equation}
\langle\Phi_{l_2}\Phi_{-l_1-l_2} \Phi_{\tilde{l}_2}
\Phi_{-\tilde{l}_1-\tilde{l}_2}\rangle_c\supset\fnlloc^2\Omega^3
\delta_{l_1,-\tilde{l}_1} C_{l_2}C_{l_1}C_{\tilde{l}_2} .
\end{equation}
Considering the effect of this term in the variance in
eq.~(\ref{exvar}), where we also take the 2-point function at
zeroth order in $\fnlloc$, we obtain:
\begin{eqnarray}
\langle\Delta\cale^2\rangle
\supset\frac{\fnlloc^2}{N^2}\sum_{l_1 l_2
\tilde{l}_2}C^2_{l_1} \propto \frac{\fnlloc^2 A \, \Npix^2}{N^2}\propto\frac{\fnlloc^2}{\ln^2\Npix} \;.
\end{eqnarray}
Thus this contribution to the relative
variance {\it does not} decrease as $1/(\Npix \ln\Npix)$, as one would have
naively expected, but there is an enhancement of $\Npix$ which
make it decrease only as $1/\ln^2\Npix$. 

There is a very
physical reason for the presence of such enhanced terms for the
case of local non-Gaussianities. As we have already said, most of
the signal for this kind of non-Gaussianities comes from squeezed
configurations, where one of the $l$s is small and the others are
large. More precisely, for the 3-point function, the signal-to-noise from all the squeezed configurations with the smallest $l$ in a given decade is roughly the same for every decade, although there are much fewer modes in a decade of low $l$. This means that the low $l$ modes are
always very important for the estimator $\cale$. Now, the point is
that there is an intrinsic variance associated with a
configuration with a certain small $l$, simply because there are
very few of those small $l$s, just $2l+1$; and this is unaffected by the fact that $\Npix$ of the survey
increases, because this just increases the $l_{\max}$ of the
experiment. Therefore the relative variance of the estimator due to these terms
decrease only logarithmically with $\Npix$, because
this is how the relative importance of the configurations with
small $l$s decreases with $\Npix$. This physical explanation
guarantees us that these enhanced terms are not present in the
case of equilateral non-Gaussianities, where the importance of the
small $l$s and of the squeezed configurations is marginal. 

It is useful to develop a quick thumb rule to understand if a
term which is a sum of product of different $C_l$s is enhanced or
not: a term will be enhanced only if at
least one $C_l$ is raised to a power larger than one. In
fact in this case the summation over the multipoles for this term will be dominated
by the lowest $l$s, so that some $l_{\rm min}$ will appear in
the denominator. This makes these terms enhanced by powers of $l_{\rm max}/l_{\rm min}$ (see also
appendix \ref{appA}).

\paragraph{The relevance of the additional terms.} Our discussion shows  that the treatment of
expressions containing $\fnlloc$ is delicate in the case of
local non-Gaussianities. The expansion parameter {\it is not} just
$\fnlloc A^{1/2}$, but there are terms which can parametrically go
as $\fnlloc A^{1/2}\Npix$, and therefore cannot be neglected. We need to understand the relevance of these terms 
both for existing limits on $\fnl$ as well as their impact on future measurements.  

After a careful calculation, we find the following expression for
the variance of $\cale$:
\begin{equation}
\langle\Delta\cale^2\rangle=\frac{1}{
4 A \Npix\ln\Npix}\left(1+\frac{8\fnlloc^2A \Npix }{\pi
\ln\Npix}+\cdots\right)\label{variance} ,
\end{equation}
where $\cdots$ represents terms suppressed by powers of $\fnlloc^2
A$ without any further $\Npix$ enhancement. This result shows an
important feature of this estimator. Imagine that we have a
series of experiments with increasing $\Npix$, and that at some
point we detect a non-null $\fnlloc$. Then, at first the variance
will decrease as $1/\left(\Npix\ln\Npix\right)$, but, after a
critical $\Npix$ which depends on the actual value of $\fnlloc$,
and which is basically, apart for logarithms, when the
signal-to-noise is of order 1, the variance will begin to decrease
very slowly as $1/\ln^2\Npix$, because of the enhanced variance
of the term proportional to $\fnlloc^2$.

In the analysis performed in \cite{Creminelli:2005hu,Komatsu:2003fd} the variance for a non-zero $\fnlloc$ was assumed to be the same as for $\fnlloc=0$, expecting that the $\fnlloc$ corrections
would have been small. In the light of the results of this
section, we see that this procedure is not always justified. However, for those analysis, we can verify that the error introduced is very small, as already numerically checked with non-Gaussian Montecarlos in \cite{Komatsu:2003fd,Komatsu:2002db}. We can quantify the error is this way: the relative correction to the variance for an
$\fnlloc$ at $n\,\sigma_0$ from the origin, where $\sigma_0$ is
the variance computed at $\fnlloc=0$,  is of order $2 n^2/(\pi\ln^2\Npix)$.
For the WMAP experiment $\ln^2\Npix\sim 35$, therefore
this correction is large for $n$ larger than $\sim 6, 7$. Therefore, if we wish to
give a $2-\sigma$ confidence interval around a certain central
value, we see that the enhanced terms will
become important for a central value around $4,5-\sigma_0$ far
from the origin,~i.e. in the case of a clear detection of a non
zero $\fnlloc$. In the analysis of
\cite{Creminelli:2005hu,Komatsu:2003fd}, the central value of
$\fnlloc$ is of the order of only one $\sigma_0$ far from the
origin. Because of this, the approximation done in
\cite{Creminelli:2005hu,Komatsu:2003fd} of considering for a
non-zero $\fnlloc$ the variance  at $\fnlloc=0$ is numerically
justified, with a small error at the percent level, well beneath the 
error coming from other sources, for example from the uncertainty in the cosmological
parameters, which gives an error on the variance of order ten
percent \cite{Creminelli:2005hu}.

Summarizing, we conclude that the enhanced terms will {\it not} be
important until there is a clear detection of a non-zero $\fnlloc$. If that
happens,  they will have to be taken into account. At that
point, the variance of the estimator $\cale$ will begin to
decrease as $1/\ln^2 \Npix$. Given the very slow convergence of $\cale$ in this regime, one is lead to wonder whether a better estimator exists.

\section{Likelihood Calculation}
For non-Gaussianities of the local type, it is easy to calculate
the full Likelihood for $\fnl$ given the data and determine to what extent  the data are able to constrain $\fnl$. This is true 
even in the high signal to noise limit where 
the previous estimator  has an increased variance. 
 
With the full Likelihood it is possible to determine what is 
the minimum variance that an
estimator of $\fnlloc$ can have, the so called Cramer-Rao bound. The bound on the variance is $\langle\partial^2{\cal L}/\partial\fnlloc^2\rangle^{-1}$, where ${\cal L}$ is minus the logarithm of the Likelihood. In \cite{Babich:2005en} it was proved that the estimator $\cale$ of
the former section, whose variance scales as $1/(\Npix\ln \Npix)$, satisfies
this bound at order zero in $\fnlloc A^{1/2}$. However, we have
just learned that this expansion in powers of $\fnlloc A^{1/2}$
 breaks down when $\fnlloc$ is detected  
because of the presence of enhanced terms. It is
therefore worth asking what happens to the Cramer-Rao bound in the
same regime, and check if there are enhanced term also in this
case. 

By the end of this section, we will see that the Likelihood
allows for an expansion in powers of $\fnlloc^2A$, without $\Npix$
enhancements, and therefore that the Cramer-Rao bound in the
presence of a non-null $\fnlloc$ is affected only marginally by
terms suppressed by powers of $\fnlloc^2A$. This will tell us that
the estimator $\cale$ of the previous section is just a bad estimator in the large signal to noise regime, and that in principle there can be estimators whose
variance in this regime scales as $1/(\Npix\ln \Npix)$.

\paragraph{Full Likelihood and Cramer-Rao bound: leading terms.}
The Likelihood function can be simply obtained inverting
eq.~(\ref{localnongaussdef}), and expressing the probability for the
Gaussian variables $g$ as a function of the temperature field
$\Phi$:
\begin{equation}
g_{\theta}=f^{-1}(\Phi_{\theta})=\Phi_{\theta}-\tilde f_{\rm NL}\left(\Phi_{\theta}^2-\sigma^2\right)
+2\tilde f_{\rm NL}^2
\Phi_{\theta}\left(\Phi_{\theta}^2-\sigma^2)\right)+\cdots .
\label{temp}
\end{equation}
The Likelihood will be a function of the parameter $\tilde f_{\rm NL}$, while we keep $\fnlloc$ to denote the true value of the non-Gaussianity parameter. The dots represents higher order terms in $\tilde f_{\rm NL}$ coming
from the inversion of the function $f(g_{\vec{\theta}})$, which
for the moment we neglect. We will come back to them shortly. In Fourier space, expression
(\ref{temp}) translates into:
\begin{equation}
g_{l}=\left(f^{-1}(\Phi)\right)_{l}=\Phi_{l}-\tilde f_{\rm NL}\left(\left(\Phi\circ\Phi\right)_{l}-\sigma^2
\Omega \delta_{l,\,0}\right)+
2\tilde f_{\rm NL}^2\left((\Phi\circ\Phi\circ\Phi)_{l}-\sigma^2 \Phi_{l}
\right)+\cdots .
\end{equation}

Starting from minus the logarithm of the probability
\begin{equation}
{\cal L}_g=\frac{1}{2}\sl1l2 C^{-1}_{l_1l_2}g_{l_1} g_{l_2} \  ,
\end{equation}
we change variable from $g_l$ to $\Phi_l$ taking into account the change in the measure:
\begin{equation}
{\cal L}=\frac{1}{2}\sl1l2 C^{-1}_{{l_1}l_2}
\left(f^{-1}(\Phi)\right)_{l_1}\left(f^{-1}(\Phi)\right)_{l_2}-{\rm
Tr}\ln\left({\rm J}\right) ,
\end{equation}
where ${\rm Tr}$ stays for trace in Fourier space, and ${\rm J}$
is the Jacobian
\begin{equation}
{\rm J}=\frac{\partial
\left(f^{-1}(\Phi)\right)_{l_1}}{\partial \Phi_{l_2}}.
\end{equation}

We can now expand to second order in $\tilde f_{\rm NL}$ to obtain:
\begin{eqnarray} \label{like}
{\cal L}&=&\frac{1}{2}\sum_{l}\left(\frac{1}{\Omega
C_l}\left(\Phi_{l}\Phi_{-l}-2\tilde f_{\rm NL}
\chi_{l}\Phi_{-l}+\tilde f_{\rm NL}^2\left(\chi_{l}\chi_{-l}+4\Phi_{l}\eta_{-l}
\right)\right)\right) \\ \nonumber &
&+2\tilde f_{\rm NL}\frac{\Npix}{\Omega}
\Phi_{l=0}-4\tilde f_{\rm NL}^2\frac{\Npix}{\Omega}\chi_{l=0}-2\tilde f_{\rm NL}^2\Npix
\sigma^2 ,
\end{eqnarray}
where we have introduced the field $\eta_{l}=(\chi\circ\Phi)_{l}$ 
\footnote{Notice that in this section, to keep the formulas as simple as 
possible, we have assumed that the average of $\Phi$ in the patch of the sky 
survey is observed. As it will become clear later, even 
if this was not the case, it would not change 
relevantly the results.}.  
Although the Likelihood contains all the information on the parameter
$\fnlloc$ one can derive from an experiment, its computation as a
function of $\tilde f_{\rm NL}$ can be very challenging in practice. All 
analysis to date have used an estimator of $\fnl$ rather than to calculate
the full Likelihood (e.g.~\cite{Komatsu:2003fd,Creminelli:2005hu}).

The statistical properties of ${\cal L}$ depend on the underlying true value $\fnl$. To make this explicit we can write
\begin{equation}
\Phi_{l}=g_{l}+\fnlloc\left( (g\circ
g)_{l}-\Omega\,\sigma^2\delta_{l,0}\right) .
\end{equation}
Plugging back in the expression for the Likelihood, we obtain:
\begin{eqnarray}
{\cal L}&=&\frac{1}{2}\sum_{l}\frac{1}{\Omega\;C_l}\left(g_{l}g_{-l}+2(\fnlloc-\tildefnlloc)
\left(\tilde\chi_{l} g_{-l}
-2\fnlloc(g_{l}\tilde\eta_{-l}-\langle
g_{l}\tilde\eta_{-l}\rangle)\right)\right.
\\ \nonumber & &\left.+\left(\fnlloc-\tildefnlloc\right)^2\left(\tilde\chi_{l}\tilde\chi_{-l}
+2g_{l}\tilde\eta_{-l}+ 2(g_{l}\tilde\eta_{-l}-\langle
g_{l}\tilde\eta_{-l}\rangle)\right)\right)\\
\nonumber & & +2\tildefnlloc\frac{\Npix}{\Omega}
g_{l=0}+\left(2\fnlloc\tilde f_{\rm NL}-4\tildefnlloc^2\right)\frac{\Npix}{\Omega}\tilde\chi_{l=0}-2 \fnlloc^2
\Npix \sigma^2 ,
\end{eqnarray}
where we have analogously defined $\tilde\chi_{l}=(g\circ g
)_{l}-\Omega\;\sigma^2 \delta_{l,0}$, and
$\tilde\eta_{l}=(\tilde\chi\circ g)_{l}$.

We can use the expression of the Likelihood we have just derived to find  the Cramer-Rao bound 
on the variance of an unbiased estimator for $\fnlloc$:
\begin{equation}
\langle\frac{\partial^2 {\cal L}}{\partial
\tildefnlloc^2}\rangle^{-1}=\left(\sum_{l}\frac{1}{\Omega\;C_l}\langle\tilde\chi_{l}\tilde\chi_{-l}
+2g_{l}\tilde\eta_{-l}\rangle \right)^{-1}=\frac{1}{8\Npix\sigma^2} \ ,
\label{cramer_rao}
\end{equation}
where we have used that:
\begin{eqnarray}
&&\sum_{l}\frac{1}{\Omega\;C_l}\langle
g_{l}\tilde\eta_{-l}\rangle= \sum_{l}\frac{1}{\Omega C_l} \langle
g_{l} \frac{1}{\Omega}\sum_{l'}g_{-l-l'}
\left(\sum_{l''}g_{l'-l''}g_{l''} -\Omega \sigma^2 \delta_{l'
,0}\right)\rangle\\ \nonumber &=& \sum_{l\,
l'l''}\frac{2}{\Omega^2 C_l} \Omega C_l \delta_{l,-l'+l''} \Omega
C_{l''}\delta_{-l-l',-l''}=2\frac{\Npix}{\Omega}\sum_{l}C_l=2\Npix\sigma^2
,
\end{eqnarray}
and analogously:
\begin{equation}
\sum_{l}\frac{1}{\Omega\;C_l}\langle\tilde\chi_{l}\tilde\chi_{-l}\rangle=4\Npix\sigma^2
.
\end{equation}
We see from eq.~(\ref{var0}) that the estimator of the last section saturates the Cramer-Rao bound for sufficiently small $\fnlloc$.

\paragraph{The expansion of the Likelihood to second order is consistent.} 
At this point, one may wonder if the
higher order terms in $\fnlloc$ might relevantly alter this result
with terms that, though suppressed by powers of $\fnlloc$, are
 enhanced by factors of $\Npix$, as in the former
section for the variance of the estimator $\cale$. It is quite
straightforward to check that this is not the case. For example, at quartic level
in $\fnlloc$ there are terms like:
\begin{eqnarray}
&&{\cal L}\supset\fnlloc^4\sum_{l}\frac{1}{\Omega\;C_l}\left( g \circ g\circ g
\right)_{l}
\left(g\circ g\circ g\right)_{-l}\\ &&\nonumber = \fnlloc^4
\frac{1}{\Omega^5}\sum_{l\;l_1l_2\tilde{l}_1\tilde{l}_2}g_{l-l_1-l_2}g_{l_1}
g_{l_2}\frac{1}{C_l}g_{-l-\tilde{l}_1-\tilde{l}_2}g_{\tilde{l}_1}g_{\tilde{l}_2} ,
\end{eqnarray}
whose expectation value contributes to the Cramer-Rao bound in
eq.~(\ref{cramer_rao}) with terms like:
\begin{equation}
\fnlloc^2\sum_{l\; l_1 l_2}\frac{1}{\Omega^2} \frac{C_{l_1}
C_{l_2}C_{l-l_1-l_2}}{C_l}=\fnlloc^2\sum_{l_1l_2l_3}\frac{1}{\Omega^2}\frac{C_{l_1}
C_{l_2}C_{l_3}}{C_{l_1+l_2+l_3}}\propto \fnlloc^2 \sigma^4 \Npix\ .
\end{equation}
We recognize this term as not being enhanced also thanks to our thumb
rule according to which a term is not enhanced if there are no $C_l$s
raised to a power larger than one. We conclude
that for these terms there is no $\Npix$ enhancement, and
therefore are suppressed by genuine powers of $\fnlloc^2 A$ with
respect to the leading terms in the Cramer-Rao bound in
eq.~(\ref{cramer_rao}). Higher order terms will appear in even
powers of $\fnlloc$, and will give similar contributions of the
form:
\begin{equation}
\fnlloc^{2n-2}\sum_{l_1l_2\dots
l_{n+1}}\frac{1}{\Omega^{n}}\frac{C_{l_1} C_{l_2}\dots
C_{l_{n+1}}}{C_{l_1+l_2+\dots + l_{n+1}}}\propto \fnlloc^{2n-2}
\sigma^{2n} \Npix \ ,
\end{equation}
so that we see there is no $\Npix$ enhancement for all these
terms. At quartic level in $\fnlloc$ there are also terms
appearing from the expansion of the Jacobian in the Likelihood, as
for example:
\begin{equation}
\fnlloc^4\frac{\Npix}{\Omega^4}\sum_{l_1l_2 l_3}g_{l_1}g_{l_2}g_{l_3}g_{-l_1
-l_2-l_3} ,
\end{equation}
whose expectation value gives subleading not-enhanced terms to the
Cramer-Rao bound of the form:
\begin{equation}
\fnlloc^2\frac{\Npix}{\Omega^2}\sum_{l_1l_2}C_{l_1}C_{l_2}=
\fnlloc^2\frac{\Npix}{\Omega^2}\left(\sum_{l}C_{l}\right)^2\propto
 \fnlloc^2 \sigma^4 \Npix ,
\end{equation}
and the same conclusion applies unaltered to the higher order
terms coming from the expansion of the Jacobian:
\begin{eqnarray}
\fnlloc^{2n-2}\frac{\Npix}{\Omega^{n}}\sum_{l_1l_3\dots
l_{2n-1}}C_{l_1}C_{l_3}\dots C_{l_{2n-1}}=\fnlloc^{2n-2}\frac{\Npix}{\Omega^{n}}\left(\sum_{l}C_{l}\right)^n\propto
 \fnlloc^{2n-2} \sigma^{2n} \Npix .
\end{eqnarray}
We conclude that these higher order terms, being not enhanced by $\Npix$, do not alter significantly the Cramer-Rao bound in
eq.~(\ref{cramer_rao}).

In order to verify the consistency of the expansion at quadratic order in $\fnl$ for the Likelihood
as well,
we need to check that the higher order terms are irrelevant not only on average, but also
on each realization. 
In appendix \ref{appA}, we show that their variance scales at most as $\Npix^2$, 
which makes their contribution to the 
Likelihood suppressed by powers of $\fnl A^{1/2}$ with respect to the contribution of the quadratic terms. We conclude that the expansion of the Likelihood 
up to quadratic order is consistent.

\paragraph{Higher order terms in the relation between $\Phi$ and $g$ are negligible.} 
Since in this section we have been very careful in keeping track
of higher order terms in $\fnlloc$, it is useful to comment  on the possibility that additional contributions come  from the presence of higher order terms in the relation between $\Phi$ and the underlying Gaussian field of eq.~(\ref{localnongaussdef}):
\begin{equation}
\Phi_{\theta}=g_{\theta}+\fnlloc (g_\theta^2 - \sigma^2) + \alpha\fnlloc^2
g_\theta (g_\theta^2-\sigma^2) + \cdots \label{localnongaussdef2}
\end{equation}
with $\alpha$ an unknown real parameter. Physically we expect these corrections to be there with $\alpha$ of order one. We need the Likelihood at second order in $\fnlloc$, so that one might worry that our results now depend on $\alpha$. This would be very strange as physically these third order terms in the expansion above are a very small correction to the, already small, second order terms. The data should not be sensitive at all to $\alpha$. One can check that this is indeed what happens. We leave the    
details of the algebra to appendix \ref{alpha}. We will find that, although $\alpha$ enters into the Likelihood at order $\fnlloc^2$,  terms containing $\alpha$ cancel on each realization up to terms suppressed by $1/\Npix$. Therefore there is no sensitivity to $\alpha$ and it can safely be set to zero. \\

In summary, we have written the Likelihood for $\fnlloc$ up to second order in $\fnlloc$, 
proving that this expansion is consistent, with the higher order terms only giving negligible 
contributions suppressed by powers of $\fnlloc A^{1/2}$. This has also allowed us to verify that 
higher order terms in $\fnlloc$ in the Likelihood do not give rise to enhanced contributions to the 
Cramer-Rao bound. 
Thus we conclude that the result in eq.~(\ref{cramer_rao}) is only corrected by terms of order 
$\fnlloc A^{1/2}$, without $\Npix$ enhancement. After what observed in the previous section, this  was not {\it a priori} guaranteed.

\section{No need to worry: simple estimators can saturate the Cramer-Rao bound
\label{secestimator}}

Now that we have determined the Cramer-Rao bound, in
this section we look for a new estimator which continues to have a
variance close to the bound even in the high signal-to-noise regime. We do this to further understand the origin of the enhanced terms and point out how a simple change in the estimator based on our intuitive understanding can make the estimator saturate the bound. 

We will start from the original estimator $\cale$ in sec.~\ref{subtleties}, and we will explicitly show
the way in which the enhanced terms cause the slow convergence of the estimator in the large signal-to-noise regime.  After this it will become easy to guess a new estimator that, 
apart from small corrections, saturates the Cramer-Rao
bound when the enhanced terms are large.

\paragraph{Explicit origin of the increased variance.}
Let us therefore start from the estimator $\cale$ in
eq.~(\ref{estimator}) and express it in terms of Gaussian variables
as:
\be
\cale = \frac{1}{N\;\Omega}\sum_{l}\frac{1}{C_l}\left(
g_l \tilde\chi_{-l}+\fnlloc
\left(\tilde\chi_{l}\tilde\chi_{-l}+2g_{l}\tilde\eta_{-l}\right)\right)\equiv\cale_0+\fnlloc
\cale_1+\cdots
\ee
where $+\cdots$ represent higher order terms in $\fnlloc$, and
where we have defined:
\begin{equation}
\cale_0=\frac{1}{N\;\Omega}\sum_{l}\frac{1}{C_l}g_l\tilde\chi_{-l} \;,
\end{equation}
\begin{equation}
\cale_1=\frac{1}{N\;\Omega}\sum_{l}\frac{1}{C_l}\left(
\tilde\chi_{l}\tilde\chi_{-l}+2g_{l}\tilde\eta_{-l}\right) .
\end{equation}
Notice that $\langle\cale_0\cale_1\rangle=0$,
$\langle\cale_0\rangle=0$, and $\langle\cale_1\rangle=1$. Therefore the variance of the estimator can be written as:
\begin{equation}
\langle\Delta\cale^2\rangle=\langle\cale^2_0\rangle+\fnlloc^2\langle\Delta\cale^2_1\rangle \;.
\end{equation}
As we discussed, the variance at zeroth order in $\fnlloc$ is 
\begin{equation}
\langle \Delta\cale^2\rangle_{\fnlloc=0}=\langle
\cale_0^2\rangle=\frac{1}{8\Npix\sigma^2}
\end{equation}
which saturates the Cramer-Rao bound in (\ref{cramer_rao}).
However, as we noted in sec.~\ref{subtleties}, the variance of
$\cale_1$ behaves like:
\begin{equation}
\langle\Delta\cale^2_1\rangle\sim\frac{2}{\pi \ln^2\Npix} \;,
\end{equation}
decreasing only logarithmically. Therefore it is going to dominate the variance of the estimator for
$\Npix/\ln\Npix \gtrsim 1/(\fnlloc^2 A)$. Apart for the logarithm,
this is when the signal-to-noise becomes of order 1.

The enhanced variance of $\cale_1$ could have been anticipated. This term contains, at leading order in $\fnlloc$, all the signal of non-Gaussianity.  We
know that for the local kind of non-Gaussianity the signal comes from
squeezed configurations with two of the three $l$s very large, and
one of the $l$s very small. The contribution to $\cale_1$ of squeezed configurations with the smallest of the $l$s within a certain decade is roughly independent of the decade, although there are very few low $l$ multipoles. {\em The value of $\cale_1$ on a given realization depends quite strongly on the particular value of the few lowest multipoles}; this explains why it converges to its average $\langle\cale_1\rangle=1$ very slowly \footnote{As shown in appendix \ref{appA}, terms in
$\cale$ of higher order in $\fnlloc$, even including a possible
contribution from terms proportional to $\alpha$, contribute to
the variance of the estimator with terms which are not enhanced by
powers of $\Npix$ more than what $\cale_1$ already is.
 Therefore they are suppressed with respect to the contribution of
$\cale_1$ by a genuine power of $\fnlloc A^{1/2}$.}. Progressively the contribution of the first decades of modes becomes negligible, so that the dependence on the particular value of the lowest multipoles goes away. However this happens only logarithmically in ${\rm N_{pix}}$ as this is the way in which the contribution from the lowest multipoles decays.

\paragraph{Improved Estimator.} 
Now that we understand better the problem of the estimator $\cale$, it is easy to find an improved estimator for the large signal-to-noise regime. We can think about the large variance of $\cale$ as coming from a ``wrong normalization". Although the estimator is clearly unbiased its value strongly depends on the amplitude of the low $l$ modes, so that if on a particular realization we have a small amplitude in the first multipoles, the value of the estimator will be small and viceversa. This effect cancels on average (that is why the estimator is unbiased) but it is the source of the large variance. Anyway this effect can clearly be corrected as we surely know the amplitude of the low $l$ modes in each particular realization: we just have to divide by a ``realization dependent" normalization. We define a new estimator $\tilde\cale$
\begin{eqnarray}
\tilde\cale&=& \frac{N \Omega}{\sum_{l}\frac{1}{C_l}
\left(\chi_{l}\chi_{-l}+2\Phi_{l}\eta_{-l}\right)} \cale =  \frac{\sum_{l}\frac{1}{C_{l}}
\Phi_{l}\chi_{-l}}{\sum_{l}\frac{1}{C_l}
\left(\chi_{l}\chi_{-l}+2\Phi_{l}\eta_{-l}\right)} = \\ &= & \fnlloc+\frac{\sum_{l}\frac{1}{C_l}\tilde g_l\chi_{-l}}
{\sum_{l}\frac{1}{C_l}
\left(\tilde\chi_{l}\tilde\chi_{-l}+2g_{l}\tilde\eta_{-l}\right)}+\cdots
=\fnlloc+\frac{\cale_0}{\cale_1}+\cdots \nonumber
\end{eqnarray}
where in the second line we have expressed everything in terms
of Gaussian variables. Neglected terms are suppressed with respect to the ones we kept by genuine powers of $\fnlloc A^{1/2}$. Neglecting these terms, the new estimator $\tilde\cale$ is unbiased:
$\langle\tilde\cale\rangle=\fnlloc$, as $\cale_0/\cale_1$ is an
odd function of the Gaussian variables $g$ and it has thus zero average. 

We can
now verify that the new estimator converges to the Cramer-Rao bound. We can write $\cale_1=1+\delta\cale_1$, where $\delta\cale_1$ is of the order
of $\langle\delta\cale_1^2\rangle^{1/2}\sim 2^{1/2}/(\pi^{1/2}
\ln\Npix)$. For large $\Npix$ we can thus expand the denominator 
\begin{equation}
\tilde\cale\simeq \fnlloc+\cale_0-\cale_0 \delta\cale_1 \;.
\end{equation}
The variance introduced by the third piece scales like
$1/(\Npix\ln^3\Npix)$, more rapidly than the
Cramer-Rao bound $\propto1/(\Npix\ln\Npix)$. After a while we are therefore
left with the variance of $\cale_0$ that, as we know, satisfies the Cramer-Rao
bound. It is worth noticing that already at the level of the WMAP experiment $\ln^2\Npix\simeq 35$, so the deviation of this estimator from the Cramer-Rao bound is already rather small. The important point is that this good behavior of $\tilde\cale$ is not spoiled when we enter in the large signal-to-noise regime.

\paragraph{The improved normalization only depends on the large scales.}
Our understanding of the enhanced variance of the estimator $\cale$ relies on the fact that always a
significant fraction of the signal is coming from low $l$ modes with a great intrinsic variance. If this is true, it better be
that the solution to this problem depends strongly on the low
$l$s. Here we therefore verify that $\cale_1$ can be written to
good approximation in terms of just the first few modes. In
appendix \ref{alpha} we have shown that
\begin{equation}
S_2=\sum_{l} \frac{1}{\Omega \, C_l}g_{l}\tilde\eta_{-l}
\end{equation}
 is fully 
correlated with the quantity
\begin{equation}
S_1=\frac{\Npix}{\Omega^2}\sum_{l} g_{l}g_{-l} \;,
\end{equation}
up to corrections ${\cal O}(1/{\rm N_{pix}})$, so that on each realization $\Delta S_2=3\Delta S_1$ with very good accuracy. In the same fashion one can prove that also the quantity
\begin{equation}
S_3=\sum_{l}\frac{1}{\Omega \; C_l}\tilde\chi_{l}\tilde\chi_{-l}
\end{equation}
 is fully correlated to $S_1$ (up to corrections ${\cal O}(1/{\rm N_{pix}})$) and that $\Delta S_3=4\Delta S_1$. This
implies that also $\cale_1$ is fully correlated with $S_1$ and
therefore, on each realization, we can write:
\begin{equation}
\cale_1=\langle\cale_1\rangle+\frac{10\Delta S_1}{N} \;.
\end{equation}
Now, the important point is that in order to compute the quantity $\Delta S_1$  on a given realization one needs, to good approximation, only the first few modes. This can be seen from
the computation of the variance of $S_1$ in Appendix \ref{alpha}:
\begin{equation}
\langle\Delta S_1^2\rangle=2{\Npix^2 \over \Omega^2} \sum_{l} C^2_l\simeq { 2\pi A^2
\Npix^2 \over \Omega} \left(\frac{1}{l^2_{\rm min}}-\frac{1}{l^2_{\rm
max}}\right)
\end{equation}
which shows that the contribution to the variance of the high $l$s
is completely irrelevant.  We therefore conclude that the value of $\cale_1$ on each realization
can be determined just by looking at the first few modes, in
agreement with our intuition.

This last remark has also relevant consequences from the
computational point of view. In fact it seems at first sight very hard to use the new estimator in the analysis of CMB data, as it contains 4-point functions and one has to deal with the complications of the spherical geometry and of the transfer function. On the other hand the dependence on only the first few modes makes the modification computationally quite light.

\paragraph{Relation to the full Likelihood calculation.} The Likelihood (\ref{like}) contains all the information on the parameter $\fnlloc$. To reconstruct it from the data we just need the coefficients of the terms linear and quadratic in $\tilde f_{\rm NL}$. {\em These two combinations of the data are sufficient statistics for $\fnlloc$}. Notice that given our discussion above it is not so complicated to analyze the full Likelihood function: we just need the same kind of terms entering in the estimator $\tilde\cale$ above. A natural question is whether one can get better constraints on $\fnlloc$  using the full Likelihood function instead of an estimator.

First of all it is straightforward to check that the maximum Likelihood estimator, which can be easily derived from eq.~(\ref{like}), has the same good properties of our improved estimator discussed above. 
It is unbiased 
up to corrections ${\cal{O}}({\rm 1/{N^{1/2}_{pix}}})$ 
and it asymptotically saturates the Cramer-Rao bound up to corrections decaying as $1/\ln^2\Npix$.    

The Cramer-Rao bound, being the average value of the second derivative of the log-Likelihood, gives the average value of error bars that one gets. The difference in using the full Likelihood is that the curvature of it changes realization by realization, as it is given by the $\tilde f_{\rm NL}^2$ term in eq.~(\ref{like}). Usually this distinction between the curvature of the Likelihood in a particular realization and its average value is irrelevant, as the difference scales like $1/\Npix$. This is not true in our case. The variance of the curvature of the Likelihood function only scales as $1/\ln^2\Npix$. This is again intuitive: given the strong dependence on the lowest multipoles, a realization with an excess of power in the low $l$s compared with the average will be more constraining than one with suppressed power on large scales. In the first case in fact it is easier to see the non-Gaussian correlation between the low $l$s and the short scale power. This difference is anyway not that large: for real experiments that have a chance of detecting $\fnl$,  $1/\ln\Npix$ is rather small. 

We reach an important conclusion.  {\em The use of our improved estimator, or equivalently the maximum Likelihood one, is equivalent to the full Likelihood of the data up to small corrections}  suppressed by $1/\ln\Npix$. This closes the door to any additional attempt to improve the limits on $\fnlloc$. 

\section{Comments on estimating $\fnlloc$ using the 4-point function}

It has been proposed in \cite{Okamoto:2002ik} and more recently in
\cite{Kogo:2006kh} that an estimator for $\fnlloc$ based on the 4-point function  has a variance which decreases as $1/\Npix^2$, so that it might be better than a 3-point function estimator for large enough $\fnl$. Of course our analysis in the previous sections shows 
that the Cramer-Rao bound scales as $1/(\Npix\ln\Npix)$, so that no estimator can do better than this. Moreover we proved that a slight modification of the 3-point function estimator makes the new estimator $\tilde\cale$ approach asymptotically the Cramer-Rao bound.
Rather than stop here and rely on the above ``theorems" we want to show explicitly in this section what goes wrong in the naive calculation of the variance of the 4-point function estimator. We will see that the $1/\Npix^2$ scaling of the variance does not hold once the signal-to-noise is larger than one and the
``enhanced" terms are taken into account. We will also show explicitly that there is
no additional information about $\fnlloc$ in the 4-point function that is not already captured by the 3-point function.

\paragraph{The variance of the proposed 4-pt estimator has enhanced terms.}
Let us begin proving that the variance of the estimator introduced
in \cite{Kogo:2006kh,Okamoto:2002ik} does not scale as
$1/\Npix^2$. The proposed estimator, analogously to the estimator $\cale$ in the 3-point function case,  is the linear combination of 4-point correlators which maximizes the signal-to-noise in the limit $\fnl \rightarrow 0$:
\begin{equation}
\cale_4=\frac{1}{N_4}\sum_{l_1l_2l_3}\frac{\langle\Phi_{l_1}\Phi_{l_2}\Phi_{l_3}
\Phi_{l_4}
\rangle_{c,1}}{C_{l_1}C_{l_2}C_{l_3}C_{l_4}}\Phi_{l_1}\Phi_{l_2}\Phi_{l_3}
\Phi_{l_4} \;,\label{fourpointest}
\end{equation}
where $N_4$ is a normalization constant which makes the estimator unbiased
$\langle\cale_4\rangle=\fnlloc^2$. The sum is restricted to momentum conserving, $l_4=-l_1-l_2-l_3$,  non-degenerate quadrilaterals. The subscript $_1$ in $\langle\Phi_{l_1}\dots\Phi_{l_4}\rangle_{c,1}$ means that the
connected 4-point function is evaluated with $\fnlloc=1$. The
variance of this estimator is:
\begin{eqnarray}
&&\langle\Delta\cale^2_4\rangle=\\
&&\nonumber
\frac{\sum_{l_1l_2l_3}\sum_{\tilde{l}_1\tilde{l}_2\tilde{l}_3}\frac{\langle\Phi_{l_1}\Phi_{l_2}\Phi_{l_3}
\Phi_{l_4}
\rangle_{c,1}}{C_{l_1}C_{l_2}C_{l_3}C_{l_4}}\frac{\langle\Phi_{\tilde{l}_1}\Phi_{\tilde{l}_2}\Phi_{\tilde{l}_3}
\Phi_{\tilde{l}_4} \rangle_{c,1}}{C_{\tilde l_1}C_{\tilde
l_2}C_{\tilde l_3}C_{\tilde
l_4}}\langle\Phi_{l_1}\Phi_{l_2}\Phi_{l_3}
\Phi_{l_4}\Phi_{\tilde{l}_1}\Phi_{\tilde{l}_2}\Phi_{\tilde{l}_3}
\Phi_{\tilde{l}_4}\rangle}{\left(\sum_{l_1l_2l_3}\frac{\langle\Phi_{l_1}\Phi_{l_2}\Phi_{l_3}
\Phi_{l_4}
\rangle^2_{c,1}}{C_{l_1}C_{l_2}C_{l_3}C_{l_4}}\right)^2}-\fnlloc^4 \;.
\end{eqnarray}
Here we are interested in the scaling with $\Npix$ of the different terms, therefore we do not keep track of the various combinatorial and numerical factors, and also of possible logarithmic corrections. Using the fact that the connected 4-point
function behaves, at leading order in $\fnlloc$, like:
\begin{equation}
\langle\Phi_{l_1}\Phi_{l_2}\Phi_{l_3} \Phi_{l_4}\rangle_c\sim
\fnlloc^2 \cdot C_{l_1}C_{l_1+l_2}C_{l_4}+{\rm symm} \;,
\end{equation}
we find that the denominator of the first term contains terms that
behave like:
\be
\sum_{l_1l_2l_3}\frac{\langle\Phi_{l_1}\Phi_{l_2}\Phi_{l_3}
\Phi_{l_4} \rangle^2_{c,1}}{C_{l_1}C_{l_2}C_{l_3}C_{l_4}}\sim
\sum_{l_1l_2l_3} \frac{C_{l_1}^2C_{l_1+l_2}^2C_{l_4}^2}{C_{l_1}C_{l_2}C_{l_3}C_{l_4}} \simeq\sum_{l_1l_2l_3}
\frac{C_{l_1}C^2_{l_1+l_2}C_{l_4}}{C_{l_2}C_{l_3}} \;.
\ee
We already met these kind of summations and we know they are dominated by $l_1+l_2\sim l_{\rm min}$ and enhanced by a factor of $\Npix$ with respect to the naive scaling. 
\begin{eqnarray}
\sum_{l_1l_2l_3}\frac{\langle\Phi_{l_1}\Phi_{l_2}\Phi_{l_3}
\Phi_{l_4} \rangle^2_{c,1}}{C_{l_1}C_{l_2}C_{l_3}C_{l_4}} \sim 
\Npix^2\sum_l C^2_l \propto\Npix^2A^2 .
\end{eqnarray}
For the numerator we have to compute the
8-point function, which in general will be the sum of the
product of four 2-point function, of two connected 3-point
functions and one 2-point function, of two connected 4-point
functions, and so on. At zeroth order in $\fnlloc$, we have only
2-point functions and we obtain something which scales as the
square root of the denominator. Therefore, if ones stops at this
level, one finds that the variance decreases as $1/\Npix^2$. This
is the result obtained in \cite{Okamoto:2002ik} and
\cite{Kogo:2006kh}. 

However, as we learned in sec.~\ref{subtleties}, there are other terms in the numerator which,
though suppressed by powers of $\fnlloc^2A$, are enhanced by
powers of $\Npix$. It turns out that the product of two 4-point
functions, of the 5-point with the 3-point one, and of the
6-point with the 2-point one, all give rise to enhanced terms with
the same scaling. For example a term with two connected 4-point functions is
\begin{eqnarray}
&&\nonumber
\sum_{l_1l_2l_3}\sum_{\tilde{l}_1\tilde{l}_2\tilde{l}_3}
\frac{\langle\Phi_{l_1}\Phi_{l_2}\Phi_{l_3} \Phi_{l_4}
\rangle_{c,1}}{C_{l_1}C_{l_2}C_{l_3}C_{l_4}}\frac{\langle\Phi_{\tilde{l}_1}\Phi_{\tilde{l}_2}\Phi_{\tilde{l}_3}
\Phi_{\tilde{l}_4} \rangle_{c,1}}{C_{\tilde l_1}C_{\tilde
l_2}C_{\tilde l_3}C_{\tilde l_4}}
\langle\Phi_{l_1}\Phi_{l_2}\Phi_{\tilde{l}_3}
\Phi_{\tilde{l}_4}\rangle_c\langle\Phi_{\tilde{l}_1}\Phi_{\tilde{l}_2}\Phi_{l_3}
\Phi_{l_4}\rangle_c
\\  &&\simeq \fnlloc^4 \sum_{l_1l_2l_3}\sum_{\tilde{l}_1\tilde{l}_2\tilde{l}_3}\frac{C_{l_1} C_{l_4} C^2_{l_1+l_2}}{C_{l_2}C_{l_3}}\frac{C_{\tilde
l_1} C_{\tilde l_4}C^2_{\tilde{l}_1+\tilde{l}_2}}{C_{\tilde l_2}C_{\tilde
l_3}} +\cdots
\end{eqnarray}
where $+\cdots$ represents terms of higher order in $\fnlloc$. The sums will be dominated by the region with
$l_1+l_2\sim \lmin$ and $\tilde{l}_1+\tilde{l}_2\sim \lmin $, so that we 
obtain
\begin{equation}
\fnlloc^4\Npix^4\big(\sum_{l}C^2_{l}\big)^2\propto\fnlloc^4A^4\Npix^4
\end{equation}
as we wanted to show. Putting together the behavior of the different terms we get the scaling of the variance of the proposed estimator up to logarithmic corrections  
\begin{equation}
\langle\Delta\cale_4^2\rangle\sim\frac{1+\fnlloc^4A^2\Npix^2}{
A^2\Npix^2} ,
\end{equation}
where we see the importance of enhanced terms at numerator. 
For comparison, it is useful to write the same schematic relation for the analogous 3-point function estimator $\cale$
we discussed in sec.~\ref{subtleties}:
\begin{equation}
\langle\Delta\cale^2\rangle\sim\frac{1+\fnlloc^2A\Npix}{ A\Npix} .
\end{equation}

We notice that the enhanced terms become parametrically important
 for both the estimators when $\fnlloc^2A\Npix\sim1$ which is, apart for logarithms, the regime
when the signal-to-noise is of order one.

\paragraph{Is there additional information in the 4-point function?}
In the limit in which the enhanced terms are negligible, we see
that the variance of $\cale_4$ is of the order of the square of
the variance of $\cale$, which means that $\cale_4$ will give
parametrically the same limits on $\fnlloc$ in this regime.
However, in the previous sections, we have shown that in the
same limit the 3-point function estimator $\cale$ saturates
the Cramer-Rao bound; therefore in this regime there is nothing which
could be added by the use of the 4-point function. This agrees
with the numerical result of \cite{Kogo:2006kh}, where it is shown
that in this regime the limit on $\fnlloc$ obtained from the
4-point function estimator is always slightly worse than the one obtained using the 3-point function.
Given the Cramer-Rao bound, we can even say something more: in this regime no improvement can be achieved from combining the two estimators.

On the other hand when the enhanced terms become important we see that
the variance of both estimators does not decrease anymore (apart for logarithmic terms). In particular there no $1/\Npix^2$ scaling for the 4-point function. As it was shown in sec.~\ref{secestimator}, in this regime one must consider ``fractional'' estimators, which are not just polynomial in the data.

\paragraph{Explicit relation between the 3-point and 4-point estimators.}
It is worth pointing out an explicit relationship between the 4-point and 3-point function estimators, to show that there is really nothing new in the 4-point estimator $\cale_4$, which is not already taken into account using $\cale$.

Let us remind once again that, for local non-Gaussianities,  the signal-to-noise of the 3-point function is concentrated on squeezed configurations, where one of the three $l$s is small, and
the other two are large and almost opposite. The 3-point function estimator $\cale$ is basically doing a weighted sum of the signal contained in all the configurations, where the weight is the
signal-to-noise ratio. Therefore, the estimator $\cale$ can be well approximated by a sum over just the squeezed
configurations:
\begin{equation}
\label{3pfapprox}
\cale\propto\sum_{L}\Phi_{L}\sum_{l}\frac{\Phi_{-L-l}\Phi_{l}}{C_l}\equiv\sum_{L}
\Phi_{L}\calk_{-L}
\end{equation}
where $L$ is a large scale multipole and $l$ is a small scale
one, and $\calk_{L}$ is defined as:
\begin{equation}
\calk_{L}=\sum_{l}\frac{\Phi_{-L-l}\Phi_{l}}{C_l} \;.
\end{equation}
In the presence of a non-zero $\fnlloc$, ${\cal K}_L$ will contain, when expressed in terms of Gaussian variables a contribution
\be
{\cal K}_L \sim 2 \fnlloc \; g_L \;.
\ee
The estimator $\cale$ correlates this contribution with the long-wave mode $\Phi_{-L}$.

Analogously in the case of the estimator $\cale_4$, as we have
seen, the signal comes from squeezed configurations, where the
four  vectors $l$s are approximately opposite in pairs. Therefore, the
estimator $\cale_4$ can be written to a good approximation as a
sum over just these configurations:
\begin{equation}
\cale_4\propto\sum_{L} C_L \calk_{L}\calk_{-L} \;.
\end{equation}
Again in the presence of a non-zero $\fnlloc$ we are correlating the non-Gaussian contribution inside each of the ${\cal K}$'s, giving an average signal $\propto \fnlloc^2$.

From this it should be clear that the two estimators are clearly not independent and that the 4-point one is less efficient because the non-Gaussian contribution must come out of both the ${\cal K}$'s,  while it is more efficient to directly correlate ${\cal K}_L$ with the mode $\Phi_{-L}$ as in eq.~(\ref{3pfapprox}).



\section{Summary}

It is perhaps unfortunate that our paper is filled with so many equations, the message however is simple. The analysis of the local type of non-Gaussianity for scale invariant perturbations is somewhat more subtle than one might have guessed: a naive $\fnl A^{1/2}$ expansion is not always appropriate. The physical origin of the effect is clear: long wavelength modes modulate the amplitude of the short wavelengths and the amplitude of this modulation produced by long wavelengths of every decade in scale is the same. Cosmic variance severely affects this long wavelengths and because their relative information contribution only decreases logarithmically with the number of pixels, one ends up with large variances for the naive $\fnl$ estimators. 

The basic point is that when one calculates the normalization of the $\fnl$ estimator one uses the average level of large scale fluctuations as opposed to the power in the individual realization one happens to have. As a result, in the limit of large signal to noise, this relatively large uncertainty in the normalization of the estimator severely enhances its variance. Fortunately one knows the amplitude of the modes in a given realization by direct measurement so it is almost trivial to fix the problem by choosing a normalization that depends on the particular realization. 

Writing down the full Likelihood one can explicitly calculate how well one should in principle be able to constrain $\fnl$ and explicitly check how the effect mentioned above comes in.  One can show that a simple modification of the naive estimator recovers all the information that the data contain and that in fact using that estimator is basically equivalent to calculating the full Likelihood, up to corrections ${\cal O}(1/\ln\Npix)$. As a result,  one is also convinced that other statistics such as the 4-point function, Minkowski functionals, wavelets, etc can at best extract as much information on $\fnlloc$ as the 3-point function. In any event they would not contribute {\it additional} information on $\fnlloc$, so once the 3-point function is measured there is nothing else to be done. 


\paragraph*{Acknowledgments} We thank Daniel Babich and Eiichiro Komatsu for useful comments. L.~S.~is supported in part by funds provided by the U.S. Department of Energy (D.O.E) under
cooperative research agreement DF-FC02-94ER40818. M.~Z.~is supported by the Packard and Sloan foundations, NSF AST-0506556 and NASA NNG05GG84G.

\appendix

\section*{Appendices}

\section{Proof that the enhancement is at most of order $\Npix$\label{appA}}

In this appendix we want to prove that the variance of the sums that appear in the Likelihood and
in the estimators scales at most as $\Npix^2$, i.e. that the possible
enhancement with respect to the naive scaling is at most of order $\Npix$. In order to do this, following the discussion in sec.~\ref{subtleties}, where we explained the thumb rule for discovering
enhanced terms, it is enough to show that in the expression in
Fourier space of the variance  there is at most one $C_l$ raised at most to the power of two.
This corresponds to an enhancement of one factor of $\Npix$, while further enhancements would require either one
$C_l$ raised to
a power larger than two, or more than one $C_l$ squared.

A good rearrangement of the various terms is obtained if we start in real
space, where the terms we are interested in can be written in the general form:
\begin{equation}
\sum_{\theta_1\theta_2}C^{-1}_{\theta_1\theta_2}\ g^M_{\theta_1} 
g^N_{\theta_2}\
, \label{term}
\end{equation}
where $M$ and $N$ are two positive integers, and $C_{\theta_i\theta_j}$ is the covariance matrix in real space. This is related to the one in Fourier space by the following relation:
\begin{equation}
C_{\theta_i\theta_j}=\frac{1}{\Omega}\sum_l C_l\; e^{i
l\cdot(\theta_i-\theta_j)} . \label{CovFour}
\end{equation}
Analogously, $C^{-1}$ can be expressed in real space as:
\begin{equation}
\label{CinvFour}
C^{-1}_{\theta_i\theta_j}=\frac{\Omega}{\Npix^2}\sum_l \frac{1}{C_l}\; e^{i
l\cdot(\theta_i-\theta_j)} .
\end{equation}

Let us compute the variance of the general term in eq.~(\ref{term}). This will
be the sum of terms of the form:
\begin{equation}
\sum_{\theta_1\theta_2\theta_3\theta_4}C^{-1}_{\theta_1\theta_2}C^{-1}_{\theta_3\theta_4}(C_{\theta_1\theta_2})^{\alpha}
(C_{\theta_1\theta_3})^{\beta}(C_{\theta_1\theta_4})^{\gamma}(C_{\theta_2\theta_3})^{\delta}
(C_{\theta_2\theta_4})^{\Sigma}(C_{\theta_3\theta_4})^{\rho}\ ,
\end{equation}
with the positive integers $\alpha,\beta,\gamma,\delta,\Sigma,\rho$ 
constrained
to
satisfy $\alpha+\beta+\gamma+\delta+\Sigma+\rho=(N+M)/2$.

We can now express each of the $C$s and $C^{-1}$s in Fourier space with the relations (\ref{CovFour}) and (\ref{CinvFour}).
After this, the summation over the
angles $\theta_1,\theta_2,\theta_3$ and $\theta_4$ becomes trivial,
each of these giving a Kronecker
delta. In particular, the summations over $\theta_1$ and $\theta_2$ give the
constraints:
\begin{equation}
l_1+l^{\alpha}_{1}+\cdots+ l^{\alpha}_{\alpha}+
l^{\beta}_{1}+\cdots+ l^{\beta}_{\beta}+l^{\gamma}_{1}+\cdots+
l^{\gamma}_{\gamma}=0\ ,
\end{equation}
\begin{equation}
-l_1-l^{\alpha}_{1}-\cdots- l^{\alpha}_{\alpha}+
l^{\delta}_{1}+\cdots+ l^{\delta}_{\delta}+l^{\Sigma}_{1}+\cdots+
l^{\Sigma}_{\Sigma}=0
\end{equation}
where $l_1$ is the $l$ associated to the Fourier transform of
$C^{-1}_{\theta_1\theta_2}$,
$l^{\alpha}_i$, with $i=1,\dots,\alpha$, are the $l$s associated to the 
Fourier
transform of
$(C_{\theta_1\theta_2})^{\alpha}$, and analogously for the other $C$s.
These two constraints can be usefully rewritten as:
\begin{equation}
\label{cons1}
l_1=-l^{\alpha}_{1}-\cdots-l^{\alpha}_{\alpha}-
l^{\beta}_{1}-\cdots-l^{\beta}_{\beta}-l^{\gamma}_{1}-\cdots-l^{\gamma}_{\gamma}\
,
\end{equation}
\begin{equation}
l^{\beta}_{1}+\cdots+l^{\beta}_{\beta}+l^{\gamma}_{1}+\cdots+l^{\gamma}_{\gamma}+l^{\delta}_{1}+\cdots+
l^{\delta}_{\delta}+l^{\Sigma}_{1}+\cdots+ l^{\Sigma}_{\Sigma}=0
\label{trivial}\ .
\end{equation}
 From the summation over $\theta_3$ and $\theta_4$, we 
obtain two analogous constraints that can be written as:
\begin{equation}
\label{cons3}
l_2=l^{\beta}_{1}+\cdots+l^{\beta}_{\beta}+
l^{\delta}_{1}+\cdots+l^{\delta}_{\delta}-l^{\rho}_{1}-\cdots-l^{\rho}_{\rho} \ ,
\end{equation}
\begin{equation}
l^{\beta}_{1}+\cdots+l^{\beta}_{\beta}+l^{\gamma}_{1}+\cdots+l^{\gamma}_{\gamma}+l^{\delta}_{1}+\cdots+
l^{\delta}_{\delta}+l^{\Sigma}_{1}+\cdots+ l^{\Sigma}_{\Sigma}=0\ ,
\end{equation}
where $l_2$ is the $l$ associated to the Fourier transform of
$C^{-1}_{\theta_3\theta_4}$. We see that the second of these constraints is
equivalent to the one in eq.~(\ref{trivial}). The presence of a redundant constraint is
a manifestation of the fact that the term we started with in eq.~(\ref{term})
was rotationally invariant.

After the summation over the angles, we are left with summations only over the $l$s:
\begin{equation}
\sum_{l_1l_2}\sum_{l^\alpha_1\dots l^\alpha_\alpha}\cdots
\sum_{l^\rho_1\cdots l^\rho_\rho}
\frac{\left(C_{l^\alpha_1}\cdots C_{l^\alpha_\alpha}\right)\cdots
\left(C_{l^\rho_1}\cdots C_{l^\rho_\rho}\right) }{C_{l_1}C_{l_2}}\ ,
\end{equation}
subject to the three independent constraints we have found. 
Notice that the $\Npix$ factors in (\ref{CinvFour}) exactly cancel with the four ones from the sums over $\theta_i$.
Now, the first and third constraints can
be used to eliminate the summations over $l_1$ and $l_2$, leaving us
with:
\begin{equation}
\sum_{l^\alpha_1\dots l^\alpha_\alpha}\cdots
\sum_{l^\rho_1\cdots l^\rho_\rho}
\frac{\left(C_{l^\alpha_1}\cdots C_{l^\alpha_\alpha}\right)\cdots
\left(C_{l^\rho_1}\cdots C_{l^\rho_\rho}\right) }{C_{l_1}
C_{l_2}}\ ,
\end{equation}
with $l_1$ and $l_2$ given by eq.~(\ref{cons1}) and (\ref{cons3}). 
The only remaining constraint is eq.~(\ref{trivial}).
After applying it, the variance is in the form such that we can quickly
apply our thumb rule, and understand its level of enhancement.
Naively given that the number of $C_l$s is $(N+M)/2-2$  and the number of summations is $(N+M)/2-1$ we get a behavior $\sim \Npix$.   
But it can happen that the last constraint makes two $C_l$s at numerator equal. In this case the sum goes as $\Npix^2$. No further enhancement is possible.

\section{Higher order corrections in the definition of local
non-Gaussianities \label{alpha}}

In Fourier space eq.~(\ref{localnongaussdef2}) reads:
\begin{equation}
\Phi_{l}=g_{l}+\fnlloc\tilde\chi_{l}+\alpha \fnlloc^2\tilde\eta_{l} \;.
\end{equation}
In this appendix we want to prove that, although $\alpha$ enters in the Likelihood at order $\fnlloc^2$, this does not imply that data are sensitive to this parameter, unless it is huge compared to the naive estimate $\alpha \sim {\cal O}(1)$.
The new term gives the following contribution to the Likelihood at order $\fnlloc^2$:
\be \label{Liklialpha}
{\cal L}_\alpha=\alpha \tilde f_{\rm NL}^2\left(\sum_{l} - \frac{1}{\Omega \,
C_l}\left(g_{l}\tilde\eta_{-l}-\langle
g_{l}\tilde\eta_{-l}\rangle\right)+3 \frac{\Npix}{\Omega}\tilde\chi_{l=0}\right) \;,
\ee
where we have neglected terms which are independent of $\tilde f_{\rm NL}$. We notice that both terms above have zero average. However this is not enough to prove that there is no relevant dependence on $\alpha$ because both terms have enhanced variance, so that they converge to zero very slowly. Their importance with respect to the other terms in the Likelihood decreases as $1/\ln^2 \Npix$. What we are now going to prove is that, although both terms have large variance, they are strongly correlated and their contributions in eq.~(\ref{Liklialpha}) cancel up to terms suppressed by $1/\Npix$. Therefore the dependence on $\alpha$ is extremely small as expected on physical grounds.

Defining:
\be
S_1=\frac{\Npix}{\Omega^2}\sum_{l}g_{l}g_{-l} \;, \qquad
S_2=\sum_{l}\frac{1}{\Omega \,C_l}g_{l}\tilde\eta_{-l} \;,
\ee
we can write eq.~(\ref{Liklialpha}) as:
\begin{equation}
{\cal L}_\alpha=\alpha \tilde f_{\rm NL}^2\left(-\Delta S_2+3\Delta S_1
\right) \;.
\end{equation}
We can now compute the
correlation functions of $S_1$ and $S_2$:
\begin{equation}
\langle S_1\rangle=\frac{\Npix}{\Omega^2}\sum_{l}\langle
g_{l}g_{-l}\rangle=\frac{\Npix}{\Omega}\sum_{l} C_l \;,
\end{equation}
\begin{equation}
\langle\Delta
S_1^2\rangle=\frac{\Npix^2}{\Omega^4}\sum_{l\,\tilde l}\langle
g_{l}g_{-l}g_{\tilde{l}}g_{-\tilde{l}}\rangle-\langle
S_1\rangle^2=2\frac{\Npix^2}{\Omega^2}\sum_{l}C^2_l \;,
\end{equation}
\begin{eqnarray}
\langle S_2\rangle=2\frac{\Npix}{\Omega}\sum_{l}C_l \;,
\end{eqnarray}
as computed before, and:
\begin{eqnarray}
\langle\Delta S_2^2\rangle&=&\sum_{l\tilde{l}}\frac{1}{\Omega
C_l}\frac{1}{\Omega C_{\tilde l}}\langle
g_{l}\tilde\eta_{-l}g_{\tilde
l}\tilde\eta_{-\tilde{l}}\rangle-\langle S_2\rangle^2\\ \nonumber
&=&\sum_{l\tilde{l}}\frac{1}{\Omega^2 C_l C_{\tilde{l}}}\langle
g_{l}\left(\frac{1}{\Omega}\sum_{l'}g_{-l-l'}\left(\frac{1}{\Omega}
\sum_{l''}g_{l'-l''}g_{l''}-\Omega\sigma^2 \delta_{l',0}
\right)\right)\\ \nonumber &&\times
g_{\tilde{l}}\left(\frac{1}{\Omega}\sum_{\tilde{l}'}g_{-\tilde{l}-\tilde{l}'}\left(\frac{1}{\Omega}
\sum_{\tilde{l}''}g_{\tilde{l}'-\tilde{l}''}g_{\tilde{l}''}-\Omega\sigma^2
\delta_{\tilde{l}',0} \right)\right)\rangle-\langle S_2\rangle^2 \;.
\end{eqnarray}
The summation is dominated by those terms which come from the
contraction of the $g_{l}$ in each $S_2$ with one of the three
$g_{l}$s contained inside the $\tilde{\eta}_{-l}$ from the same
$S_2$ term. These are enhanced by a factor of $\Npix$ with respect to the other contributions. Neglecting subleading terms we obtain:
\begin{eqnarray}
\langle\Delta S_2^2\rangle=18
\frac{\Npix^2}{\Omega^2}\sum_{l}C^2_l \; .
\end{eqnarray}
Finally for $\langle\Delta S_1\Delta S_2\rangle=\langle S_1 S_2
\rangle-\langle S_1\rangle\langle S_2\rangle$, we obtain:
\be
\langle\Delta S_1\Delta S_2\rangle = 6\frac{\Npix^2}{\Omega^2}\sum_{l}C^2_l \;,
\ee
again keeping only leading terms.
We see that $S_1$ and $S_2$ are fully correlated, up to correction ${\cal O}(1/{\rm N_{pix}})$:
\begin{equation}
\frac{\langle\Delta S_1 \Delta S_2\rangle}{\left(\langle\Delta
S^2_1\rangle \langle\Delta S^2_2\rangle\right)^{1/2}}=
\frac{6\frac{\Npix^2}{\Omega^2}\sum_{l}C^2_l}{\left(2\frac{\Npix^2}{\Omega^2}\sum_{l}C^2_l\times
18\frac{\Npix^2}{\Omega^2}\sum_{l}C^2_l \right)^{1/2}}=1\;.
\end{equation}
On each realization we have:
\begin{equation}
\Delta S_2=\left(\frac{\langle\Delta S^2_2 \rangle}{\langle\Delta
S^2_1 \rangle}\right)^{1/2}\Delta S_1=3\Delta S_1\;,
\end{equation}
and therefore ${\cal L}_\alpha=0$ up terms suppressed by $1/{\rm N_{pix}}$.

We conclude that terms which depend on $\alpha$ are negligible in the Likelihood.

\end{document}